\documentclass{article}
\usepackage{spconf,amsmath,graphicx}

\usepackage[utf8]{inputenc} 
\usepackage[T1]{fontenc}    
\usepackage{hyperref}       
\usepackage{url}            
\usepackage{booktabs}       
\usepackage{amsfonts}       
\usepackage{nicefrac}       
\usepackage{microtype}      
\usepackage{xcolor}         
\usepackage{float}

\usepackage{bm}
\usepackage{tikz}
\usepackage{pgfplots}
\usepackage{subcaption}
\usepackage{glossaries}
\usepackage{placeins}
\usepackage{array}
\usepackage{multirow}

\pgfplotsset{compat=1.16}
\usetikzlibrary{pgfplots.colormaps}
\usetikzlibrary{arrows.meta}
\usetikzlibrary{3d,arrows,automata,backgrounds,calc,calendar,chains,decorations,decorations.footprints,decorations.fractals,decorations.markings,decorations.pathmorphing,decorations.pathreplacing,decorations.shapes,decorations.text,er,fadings,fit,folding,matrix,mindmap,patterns,petri,plothandlers,plotmarks,positioning,scopes,shadows,shapes.arrows,shapes.callouts,shapes,shapes.gates.logic.IEC,shapes.gates.logic.US,shapes.geometric,shapes.misc,shapes.multipart,shapes.symbols,through,topaths,trees,pgfplots.groupplots}
\usepackage{tkz-kiviat-new}
\usepgfplotslibrary{statistics}

\title{TEST-TIME ADAPTATION FOR SPEECH ENHANCEMENT VIA MASK POLARIZATION}
\name{Tobias Raichle, Erfan Amini, Bin Yang}
\address{University of Stuttgart, Institute of Signal Processing and System Theory, Stuttgart, Germany}
\glsdisablehyper
\newacronym{rir}{RIR}{room impulse response}
\newacronym{se}{SE}{speech enhancement}
\newacronym{tta}{TTA}{test-time adaptation}
\newacronym{ttt}{TTT}{test-time training}
\newacronym{uda}{UDA}{unsupervised domain adaptation}
\newacronym{asr}{ASR}{automatic speech recognition}
\newacronym{rtf}{RTF}{real-time factor}

\newacronym{ssra}{SSRA}{self-supvervised representation based adaptation}
\newacronym{pfpl}{PFPL}{phone-fortified perceptual-loss}
\newacronym{laden}{LaDen}{latent denoising}
\newacronym{mpol}{MPol}{mask polarization}
\newacronym{diet}{DIET}{domain invariant embedding transformation}
\newacronym{tent}{tent}{test-time entropy minimization}

\newacronym{ss}{SS}{spectral subtraction}
\newacronym{am}{AM}{amplitude masking}

\newacronym{mse}{MSE}{mean squared error}
\newacronym{tf}{TF}{time-frequency}
\newacronym{pesq}{PESQ}{perceptual evaluation of speech quality}
\newacronym{sisdr}{SI-SDR}{scale-invariant signal-to-distortion ratio}
\glsunset{sisdr}
\newacronym{ssnr}{SSNR}{segmental signal-to-noise ratio}
\glsunset{ssnr}

\newacronym{vbd}{VBD}{VoiceBank+DEMAND}
\newacronym{vbw}{VBW}{VoiceBank+WHAM!}
\newacronym{dns}{DNS}{deep noise suppression}

\begin{document}
\definecolor{mittelblau}{RGB}{0, 126, 198}
\definecolor{violettblau}{cmyk}{0.9, 0.6, 0, 0}
\definecolor{rot}{RGB}{238, 28 35}
\definecolor{apfelgruen}{RGB}{140, 198, 62}
\definecolor{gelb}{RGB}{255, 229, 0}
\definecolor{orange}{RGB}{244, 111, 33}
\definecolor{pink}{RGB}{237, 0, 140}
\definecolor{lila}{RGB}{128, 10, 145}
\definecolor{cyan}{RGB}{58, 252, 252}
\definecolor{moosgruen}{RGB}{53, 147, 134}
\definecolor{bordeaux}{RGB}{148, 23, 112}
\definecolor{kastanie}{RGB}{148, 23, 81}
\definecolor{lachs}{RGB}{255, 126, 121}
\definecolor{turkis}{HTML}{B3ECEC}
\definecolor{lavendel}{RGB}{215, 131, 254}
\definecolor{hellgrau}{RGB}{224, 224, 224}
\definecolor{mittelgrau}{RGB}{128, 128, 128}
\definecolor{dunkelgrau}{RGB}{80,80,80}
\definecolor{anthrazit}{RGB}{19, 31, 31}

\pgfplotsset{bar_1_style/.style={color=mittelblau, fill=mittelblau!50}}
\pgfplotsset{bar_2_style/.style={color=orange, fill=orange!50}}
\pgfplotsset{bar_3_style/.style={color=pink, fill=pink!50}}
\pgfplotsset{bar_4_style/.style={color=lila, fill=lila!50}}
\pgfplotsset{bar_5_style/.style={color=lavendel, fill=lavendel!50}}

\pgfplotscreateplotcyclelist{bar_default}{%
bar_1_style\\%
bar_2_style\\%
bar_3_style\\%
bar_4_style\\%
bar_5_style\\%
}

\newcommand{\PreserveBackslash}[1]{\let\temp=\\#1\let\\=\temp}
\newcommand{\newcol}{\vfill\pagebreak}

\newcolumntype{C}[1]{>{\PreserveBackslash\centering}p{#1}}

\tikzset{arrow/.style={
->,
>={Stealth[inset=0pt, angle=30:5pt]},
}}

\tikzset{arrow_large/.style={
->,
>={Stealth[inset=0pt, angle=30:12pt]},
}}

\tikzset{line_back/.style={
			color=lila,
		}}

\tikzset{arrow_back/.style={
<-,
>={Stealth[inset=0pt, angle=35:4pt]},
color=lila,
}}

\newlength{\wcol}

\newcommand{\matr}[1]{\mathbf{#1}}
\newcommand{\vect}[1]{\bm{#1}}

\newcommand{\cmark}{\ding{51}}%
\newcommand{\xmark}{\ding{55}}%
\newcommand{\glsi}[1]{\glsunset{#1}\gls{#1} (\acrlong{#1})}
\newcommand{\Glsi}[1]{\glsunset{#1}\Gls{#1} (\acrlong{#1})}

\newcommand{\dns}[1]{\(\text{DNS}_{\text{#1}}\)}
\newcommand{\avg}[1]{\(\overline{\text{#1}}\)}

\newcommand{\todo}[1]{{\color{red}TODO: }#1}

\newlength{\corner}
\newlength{\scorner}

\setlength\belowdisplayskip{6pt}%

\tikzstyle{frozen} = [double, thick, draw=turkis, fill=turkis!50, text=turkis!330, rounded corners=\scorner]
\tikzstyle{truth} = [draw=apfelgruen, fill=apfelgruen!10, text=apfelgruen!120, rounded corners=\scorner]
\tikzstyle{trainable} = [draw=lila, fill=lila!10, text=lila!80!black, semithick, rounded corners=\scorner]
\tikzstyle{norm} = [draw=orange, fill=orange!10, text=orange!90!black, semithick]
\tikzstyle{signal} = [draw=mittelgrau, text=mittelgrau!90!black, semithick, rounded corners=\corner, inner sep=0pt]
\tikzstyle{dsp} = [draw=moosgruen, fill=moosgruen!20, text=moosgruen!90!black, semithick, rounded corners=\scorner]

\pgfdeclarepatternformonly{south west lines}{\pgfqpoint{-0pt}{-0pt}}{\pgfqpoint{3pt}{3pt}}{\pgfqpoint{3pt}{3pt}}{
	\pgfsetlinewidth{0.4pt}
	\pgfpathmoveto{\pgfqpoint{0pt}{0pt}}
	\pgfpathlineto{\pgfqpoint{3pt}{3pt}}
	\pgfpathmoveto{\pgfqpoint{2.8pt}{-.2pt}}
	\pgfpathlineto{\pgfqpoint{3.2pt}{.2pt}}
	\pgfpathmoveto{\pgfqpoint{-.2pt}{2.8pt}}
	\pgfpathlineto{\pgfqpoint{.2pt}{3.2pt}}
	\pgfusepath{stroke}}

\pgfdeclarepatternformonly{south east lines}{\pgfqpoint{-0pt}{-0pt}}{\pgfqpoint{3pt}{3pt}}{\pgfqpoint{3pt}{3pt}}{
	\pgfsetlinewidth{0.4pt}
	\pgfpathmoveto{\pgfqpoint{0pt}{3pt}}
	\pgfpathlineto{\pgfqpoint{3pt}{0pt}}
	\pgfpathmoveto{\pgfqpoint{.2pt}{-.2pt}}
	\pgfpathlineto{\pgfqpoint{-.2pt}{.2pt}}
	\pgfpathmoveto{\pgfqpoint{3.2pt}{2.8pt}}
	\pgfpathlineto{\pgfqpoint{2.8pt}{3.2pt}}
	\pgfusepath{stroke}}


%
\maketitle
\begin{abstract}
	Adapting \gls{se} models to unseen environments is crucial for practical deployments, yet \gls{tta} for SE remains largely under-explored due to a lack of understanding of how \gls{se} models degrade under domain shifts.
	We observe that mask-based \gls{se} models lose confidence under domain shifts, with predicted masks becoming flattened and losing decisive speech preservation and noise suppression.
	Based on this insight, we propose \gls{mpol}, a lightweight \gls{tta} method that restores mask bimodality through distribution comparison using the Wasserstein distance.
	\Gls{mpol} requires no additional parameters beyond the trained model, making it suitable for resource-constrained edge deployments.
	Experimental results across diverse domain shifts and architectures demonstrate that \gls{mpol} achieves very consistent gains that are competitive with significantly more complex approaches.
\end{abstract}
\renewcommand{\thefootnote}{}
\footnotetext{© 2026 IEEE. Personal use of this material is permitted. Permission from IEEE must be obtained for all other uses, in any current or future media, including reprinting/republishing this material for advertising or promotional purposes, creating new collective works, for resale or redistribution to servers or lists, or reuse of any copyrighted component of this work in other works.}
\renewcommand{\thefootnote}{\arabic{footnote}}

\glsresetall
\section{Introduction}
\label{sec:intro}
By leveraging large labeled datasets to learn the complex structure of speech, deep learning based~\gls{se} has revolutionized the field.
However, these methods often suffer from performance degradation when deployed in environments that differ from their training conditions \cite{laden}.
As practical \gls{se} systems must handle diverse and time-varying acoustic environments, speaker characteristics and noise types for which no labeled data is available, robust unsupervised adaptation methods are essential for reliable deployments.
While \gls{uda} methods exist for \gls{se}, they typically require access to source data, limiting their practical applicability.
\par
\Gls{tta} has emerged as a promising solution to this challenge.
As opposed to \gls{uda}, \gls{tta} requires only the trained source model and unlabeled target data, eliminating privacy concerns associated with sharing source data and reducing storage requirements~\cite{tent}.
Crucially, \gls{tta} performs online adaptation, i.e., simultaneous to inference, making it suitable for real-world and real-time deployments that do not allow for the latency of a separate adaptation phase.
However, existing \gls{tta} methods for \gls{se} rely on additional components either via an additional encoder~\cite{laden} or a teacher-student framework~\cite{remixit}, limiting the applicability on resource-constrained deployments.
To develop a parameter-efficient alternative, we investigated how \gls{se} models degrade under domain shifts.
\par
\begin{figure}[h]
	\centering
	\resizebox{0.9\columnwidth}{!}{\input{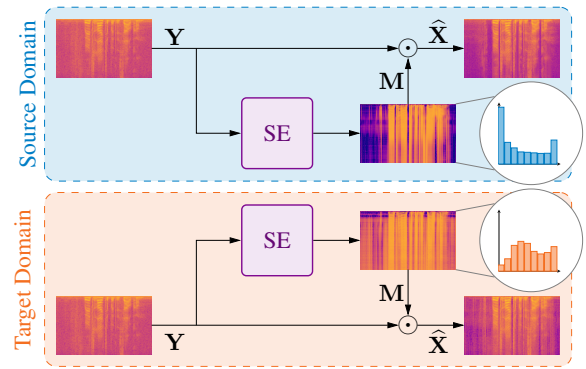}}
	\caption{Overview of \glsi{mpol}. The mask histogram flattens under domain shifts.}
	\label{fig:overview}
\end{figure}

We observe that under domain shifts mask-based \gls{se} models exhibit a phenomenon analogous to confidence loss in classification.
In their source domain, these models predict distinctly bimodal masks that effectively separate speech from noise.
However, under domain shifts, their predicted masks lose this bimodal distribution and become flatter (cf. Fig.~\ref{fig:overview}), directly impacting enhancement quality.
\par
Based on this insight, we propose \gls{mpol}, a lightweight \gls{tta} method that aims at restoring the bimodality under domain shifts by comparing the predicted mask to a computed reference via the Wasserstein distance.
\newline
The main contributions of this work are summarized as follows:
\begin{itemize}
	\item We empirically demonstrate that SE models suffer from reduced mask confidence under domain shifts.
	\item We propose \gls{mpol}, a lightweight \gls{tta} method for mask-based SE models requiring no additional components beyond the trained model.
	\item We show that \gls{mpol} achieves competitive performance with significantly more complex methods across diverse domain shifts, making it the state-of-the-art low complexity \gls{tta} method for \gls{se}.
\end{itemize}

\subsection{Problem Formulation}
Given a noisy speech recording
\begin{align}
	\vect{y} = \vect{x} + \vect{n},
\end{align}
where \(\vect{x}\) and \(\vect{n}\) denote the clean speech and noise signals respectively, the task of \gls{se} is to recover an estimate \(\hat{\vect{x}}\) of the clean speech signal~\cite{speech_enhancement}.
Generally, corruptions can include a wide range of effects including reverberation, limited bandwidth or clipping.
However, this work only considers additive noise and leaves other corruptions for future works.
A common class of \gls{se} architectures works by estimating a mask \(\matr{M}\in \mathbb{R}^{T\times F}\) in the \gls{tf} domain to suppress noise components in the noisy magnitude spectrogram \(\matr{Y}\in \mathbb{R}^{T\times F}\) (cf. Fig~\ref{fig:overview}).
The enhanced magnitude spectrogram is obtained by
\begin{align}
	\label{eq:mask}
	\widehat{\matr{X}} = \matr{M}\odot \matr{Y}=\matr{M}\odot(\matr{X}+\matr{N}),
\end{align}
where \(\odot\) denotes element-wise multiplication.
\vspace{-0.5em}
\section{Related Work}
\label{sec:related}
As \gls{se} models frequently encounter unseen target domains where labeled data is unavailable, several previous works have explored applying \gls{uda} to \gls{se}.
The most common approach identifies similar source sample to use as pseudo-labels~\cite{dotn,ssra}, but requires access to source data, defying the \gls{tta} paradigm.
\par
RemixIT~\cite{remixit} performs \gls{tta} by using a teacher model to generate pseudo-targets through remixing estimated speech and noise components, then adapting a student on this weakly-labeled data.
\Gls{laden}~\cite{laden} computes pseudo-labels by mapping representations of noisy speech to clean representations in the latent space of a large pre-trained encoder via a linear approximation.
As it requires the CNN encoder of a WavLM~\cite{wavlm} model, \gls{laden} introduces 4M additional parameters, limiting its practical applicability.
\par
Whereas \gls{tta} for \gls{se} is a relatively new area of research, \gls{tta} for classification has been more thoroughly explored, leading to better understanding of how classification models degrade under domain shifts.
Based on the observation that classification models lose prediction confidence under domain shifts, a common \gls{tta} approach is to minimize the prediction entropy~\cite{tent}.
However, transferring these insights to \gls{se} is not trivial due to the inherently different regression task.

\vspace{-0.5em}
\section{Methodology}
\label{sec:methods}
To address this gap, we investigate whether \gls{se} models exhibit analogous confidence degradation under domain shifts and propose \acrfull{mpol}, a lightweight \gls{tta} method that adapts \gls{se} models by restoring ideal \gls{tf} mask characteristics.
\par
\begin{figure}[h]
	\centering
	\resizebox{0.8\columnwidth}{!}{\input{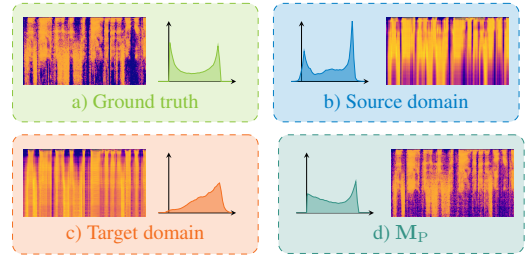}}
	\caption{Comparison of masks and their histograms.}
	\vspace{-1em}
	\label{fig:mask_comp}
\end{figure}

We observe that mask-based \gls{se} models exhibit a fundamental change in their prediction characteristic under domain shifts.
In their source domain, these models predict distinctly bimodal masks \(\matr{M}\) (cf. Fig.~\hyperref[fig:mask_comp]{2b}) that accurately preserve speech (\(\matr{M}_{ij}\approx1\)) and suppress noise (\(\matr{M}_{ij}\approx0\)).
This bimodality is consistent with the ideal ground truth mask (cf. Fig.~\hyperref[fig:mask_comp]{2a}).
However, under domain shifts, predicted masks lose this bimodal structure and become flattened, with values concentrating around an intermediate range (cf. Fig.~\hyperref[fig:mask_comp]{2c}).
\par
Deviating from the bimodal structure under domain shifts suggests that models lose confidence in their separation decisions, leading to overly conservative intermediate mask values.
This phenomenon provides a natural adaptation signal, as restoring the bimodal characteristic should improve enhancement quality by encouraging more decisive speech-noise separation.

\vspace{-0.5em}
\subsection{Mask Polarization}
Based on this insight, we propose \acrfull{mpol}, which restores mask bimodality by comparing the distribution of predicted masks with a computed reference that exhibits the desired bimodal structure.
\par
Similarly to the spectral subtraction principle~\cite{speech_enhancement}, we estimate the noise spectrum \(\widehat{\matr{N}}\) by averaging time bins of \(\matr{Y}\) with no voice activity, identified via the \(k=32\) time bins with the lowest signal power.
We compute a reference mask
\begin{align}
	\matr{M}_{\mathrm{P}}=\frac{\widehat{\matr{X}}}{\widehat{\matr{X}} + \widehat{\matr{N}}},
\end{align}
where \(\widehat{\matr{X}}\) is the enhanced \gls{tf} magnitude and the division is computed element-wise.
While \(\widehat{\matr{X}}\) comes from the model's own prediction, the reference mask \(\matr{M}_{\mathrm{P}}\) exhibits more reliable bimodality than \(\matr{M}\).
Given Eq.~\ref{eq:mask}, \(\matr{M}_{\mathrm{P}}\) can be phrased as
\begin{align}
	\matr{M}_{\mathrm{P}}=\frac{\matr{M}\odot(\matr{X} + \matr{N})}{\matr{M}\odot(\matr{X} + \matr{N}) + \widehat{\matr{N}}}
	\approx \frac{\matr{M}\odot(\matr{X}+\matr{N})}{\matr{M}\odot\matr{X} + (\matr{M} + \matr{1})\odot\matr{N}},
\end{align}
assuming \(\widehat{\matr{N}}\approx\matr{N}\).
If the \gls{tf} bin \(ij\) contains speech, we will assume that \(\matr{X}_{ij}\gg \matr{N}_{ij}\) and therefore \({\matr{M}_{\mathrm{P},ij}\approx \frac{\matr{M}_{ij}\matr{X}_{ij}}{\matr{M}_{ij}\matr{X}_{ij}}=1}\), i.e., the unwanted attenuation \(\matr{M}_{ij}\) is canceled out.
On the other hand, if the bin does not contain speech, \(\matr{X}_{ij} \ll \matr{N}_{ij}\) means \({\matr{M}_{\mathrm{P},ij}\approx\frac{\matr{M}_{ij}}{1+\matr{M}_{ij}}\leq \matr{M}_{ij}}\).
Consequently, \(\matr{M}_{\mathrm{P}}\) is more bimodal than the predicted mask \(\matr{M}\), maintaining the decisive separation that the original mask \(\matr{M}\) loses under domain shifts.
This rationale holds even at \(\mathrm{SNR}\approx 0\), due to the spectral concentration of speech.
\par
As illustrated in Fig~\ref{fig:mask_comp}, the computed mask \(\matr{M}_{\mathrm{P}}\) (Fig.~\hyperref[fig:mask_comp]{2d}) approximates the ideal mask's U-shaped distribution (Fig.~\hyperref[fig:mask_comp]{2a}), contrasting sharply with the source model's flattened distribution (Fig.~\hyperref[fig:mask_comp]{2c}).
Furthermore, the computed masks adhere to the signal's individual speech-to-noise characteristics, instead of enforcing a single standard with entropy-like losses~\cite{tent}.
\par
Due to the simplistic estimation of \(\widehat{\matr{N}}\) that does not account for non-stationarity, \(\matr{M}_{\mathrm{P}}\) cannot serve as a direct, point-wise pseudo-label for \(\matr{M}\).
Instead, we compare only the empirical distributions of the predicted mask \(\matr{M}\) and the reference mask \(\matr{M}_{\mathrm{P}}\) to leverage \(\matr{M}_{\mathrm{P}}\)'s improved bimodality.
This is achieved using the Wasserstein distance between the distributions of the mask entries, which is defined as the average difference of the sorted entries of the two masks
\begin{align}
	\mathcal{L}_{\mathrm{W}}(\matr{M}, \matr{M}_{\mathrm{P}}) = \frac1{T\cdot F} \sum_{i=1}^{T\cdot F} \vert \operatorname{vec}(\matr{M})_{(i)} - \operatorname{vec}(\matr{M}_{\mathrm{P}})_{(i)}\vert,
\end{align}
where \(\operatorname{vec}(\matr{M})_{(i)}\) corresponds to the \(i\)-th largest entry of \(\matr{M}\).
This distribution-based approach is robust to imperfect references while encouraging the desired bimodal structure.
\par
To further discourage negative mask values frequently observed under domain shifts (cf. Fig.~\ref{fig:mask_comp}), we add a penalty term
\begin{align}
	\mathcal{L}_{\mathrm{S}}(\matr{M}) = \sum_{i,j} \vert m_{ij} \vert \cdot \mathbb{I}_{m_{ij} < 0}.
\end{align}
\par
The complete \gls{mpol} loss combines both terms
\begin{align}
	\mathcal{L} = \mathcal{L}_{\mathrm{W}} + \lambda \mathcal{L}_{\mathrm{S}},
\end{align}
with \(\lambda=0.1\).
\par
To improve adaptation stability, we employ continual weight ensembling \({\theta_t \leftarrow \beta \theta_t + (1 - \beta)\theta_{0}}\) between the adapted weights \(\theta_t\) at adaptation step \(t\) and the source weights \(\theta_{0}\)~\cite{roid}, where \(\beta=0.8\) is a smoothing factor.
\cite{laden}
\section{Experiments}
\label{sec:exp}
\subsection{Datasets}
All models were trained on the source dataset EARS-WHAM! (EARS-W)~\cite{ears, wham} and evaluated on the 9 target datasets proposed by~\cite{laden} to cover a wide range of domain shifts.
The EARS-DEMAND (EARS-D)~\cite{ears,demand} dataset covers domain shifts in only the noisy environment.
Analogously, VoiceBank-WHAM! (VBW)~\cite{voicebank,wham} represents a shift only in the speech component.
To explore shifts in both components, the VoiceBank-DEMAND (VBD)~\cite{voicebank,demand,vbd} dataset and the DNS~\cite{dns_2022} dataset in the languages English, German, Italian, Russian, Spanish, and French are used.
In all domains, the entire testset is used for adaptation.

\subsection{Models}
\Gls{mpol} targets \gls{tf}-magnitude masking architectures, encompassing a significant portion of current time-frequency-domain \gls{se} models.
Following~\cite{laden}, we evaluate \gls{mpol} on two representative \gls{se} architectures to assess the generality of the mask confidence observation.
The \acrshort{am} model~\cite{laden} is a sequential \gls{am}-based architecture that is trained with a MSE error to represent simple, reconstruction-focused approaches common in resource-constrained deployments.
To represent the current state-of-the-art of \gls{se}, CMGAN~\cite{cmgan} employs a MetricGAN~\cite{metric_gan} based loss focusing on perceptual performance and additionally implements a complex additive term to enable phase correction.

\subsection{Experimental Details}
As is common in \gls{se}, we put an emphasis on perceptual performance, i.e., how natural the enhanced speech sounds to humans.
To objectively evaluate the performance of our method, we use the \glsi{pesq} metric~\cite{pesq} and the CSIG, CBAK and COVL metrics proposed by~\cite{mos}.
As these metrics can lead to misleading results~\cite{pesqetarian}, we also include the signal-level metrics \glsi{ssnr} and \glsi{sisdr}~\cite{speech_enhancement}.
\par
Besides the unadapted source model baseline, RemixIT and \gls{laden} are used as references to assess \gls{mpol}'s significance within the field of \gls{tta} for \gls{se}.
\par
In all experiments the AdamW optimizer~\cite{adam_w} was used with a learning rate of \(\alpha=5\cdot10^{-4}\).
We adapt only the normalization and output layers of each model, resulting in 8\% and 0.5\% of the parameters of the \gls{am} and CMGAN architectures being adapted, respectively.
To assess statistical significance, the key experiments were repeated 10 times.
\par
On a Nvidia A6000 GPU, the unadapted \gls{am} model achieves a \gls{rtf} of 0.005 on VBD.
Whereas RemixIT and \gls{laden} achieve \glspl{rtf} of 0.011 and 0.068 respectively, \gls{mpol}'s reduced overhead allows an \gls{rtf} of 0.007, representing a roughly tenfold improvement over \gls{laden}.
\section{Results}
\label{sec:res}
\subsection{Results Analysis}
\begin{figure}[h]
	\centering
	\resizebox{0.75\columnwidth}{!}{\begin{tikzpicture}
	\tkzKiviatDiagram[scale=0.9,label distance=0.0cm,
		gap=1.0,
		ymin=-0.15,
		ymax=0.15,
		label space=3.75cm,
        lattice=3]{{~~~~~~EARS-D},VBD,VBW,\dns{EN},{\dns{GE}\hspace{2em}~},{\dns{IT}\hspace{2em}~},\dns{RU},\dns{SP},\dns{FR}}
	\tkzKiviatLine[thick,color=mittelblau,mark=none,opacity=.5](
	0.0,
	0.0,
	0.0,
	0.0,
	0.0,
	0.0,
	0.0,
	0.0,
	0.0
	)
	\tkzKiviatLineError[thick,color=orange,opacity=.5](
	-0.056/0.023,
	-0.021/0.010,
	0.003/0.009,
	0.032/0.020,
	0.054/0.023,
	0.036/0.018,
	0.004/0.014,
	0.029/0.024,
	0.037/0.017
	)
	\tkzKiviatLineError[thick,mark size=4pt,color =lila](
	-0.053/0.006,
	0.130/0.006,
	0.027/0.011,
	0.097/0.010,
	0.129/0.016,
	0.083/0.013,
	0.048/0.007,
	0.112/0.013,
	0.130/0.008
	)
	\tkzKiviatLineError[thick,mark size=4pt,color=pink](
	0.023/0.002,
	0.055/0.002,
	0.062/0.001,
	0.044/0.003,
	0.071/0.005,
	0.057/0.003,
	0.039/0.008,
	0.070/0.005,
	0.060/0.003
	)
	\tkzKiviatGrad[prefix=,unity=1,suffix=](0)
\end{tikzpicture}}
	\resizebox{0.75\columnwidth}{!}{\begin{tikzpicture}
	\begin{axis}[%
			hide axis,
			xmin=10,
			xmax=50,
			ymin=0,
			ymax=0.4,
			legend style={font=\small, legend columns=-1},
			legend cell align={left},
		]
		\addlegendimage{mittelblau, thick}
		\addlegendentry{Source \ }
		\addlegendimage{orange, thick}
		\addlegendentry{RemixIT \ }
		\addlegendimage{lila, thick}
		\addlegendentry{LaDen}
		\addlegendimage{pink, thick}
		\addlegendentry{MPol}
	\end{axis}
\end{tikzpicture}}
	\caption{\(\Delta\)PESQ results relative to the source performance using the AM architecture (\(\mu\pm2\sigma\)).}
	\vspace{-1em}
	\label{fig:am_pesq_spider}
\end{figure}
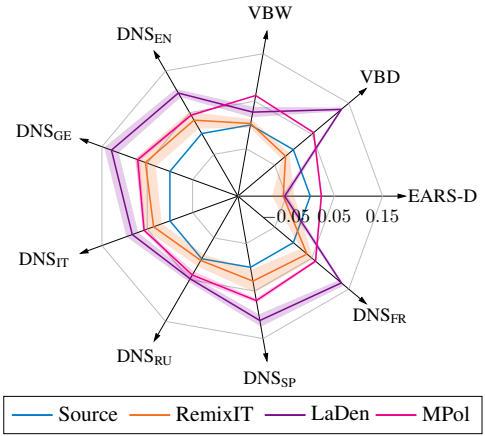

\begin{table*}[t]
	\vspace{-1em}
	\centering
	\caption{Results averaged over the datasets (\(\mu\pm2\sigma\)). SSNR and SI-SDR in dB.}
	\label{tab:avg_comp}
	\begin{tabular}{llcccccc}
	\toprule
	                                       &         & \avg{PESQ} \(\uparrow\)  & \avg{CSIG} \(\uparrow\)  & \avg{CBAK} \(\uparrow\)  & \avg{COVL} \(\uparrow\)  & \avg{SSNR} \(\uparrow\) & \avg{SI-SDR} \(\uparrow\) \\
	\midrule
	\multirow{4}{*}{\rotatebox{90}{AM}}    & Source  & 2.05                     & 3.07                     & 2.78                     & 2.52                     & 7.42                    & 12.28                     \\
	                                       & RemixIT & 2.06\(\pm\).006          & 3.10\(\pm\).007          & 2.80\(\pm\).005          & 2.54\(\pm\).006          & \textbf{7.48}\(\pm\).03 & \textbf{12.47}\(\pm\).04  \\
	                                       & LaDen   & \textbf{2.13}\(\pm\).005 & 3.13\(\pm\).007          & 2.80\(\pm\).005          & 2.59\(\pm\).006          & 7.01\(\pm\).04          & 12.33\(\pm\).04           \\
	                                       & MPol    & 2.10\(\pm\).001          & \textbf{3.17}\(\pm\).001 & \textbf{2.82}\(\pm\).001 & \textbf{2.60}\(\pm\).001 & 7.40\(\pm\).01         & 12.35\(\pm\).01          \\
	\midrule
	\multirow{4}{*}{\rotatebox{90}{CMGAN}} & Source  & 2.60                     & 3.75                     & 3.02                     & 3.15                     & 6.00                    & 11.32                     \\
	                                       & RemixIT & 2.60\(\pm\).006          & 3.77\(\pm\).006          & 3.03\(\pm\).007          & 3.17\(\pm\).007          & 5.92\(\pm\).07          & 11.52\(\pm\).07           \\
	                                       & LaDen   & 2.62\(\pm\).002          & \textbf{3.81}\(\pm\).002 & \textbf{3.07}\(\pm\).002 & \textbf{3.20}\(\pm\).002 & \textbf{6.31}\(\pm\).03 & \textbf{12.09}\(\pm\).02  \\
	                                       & MPol    & \textbf{2.64}\(\pm\).002 & 3.78\(\pm\).002          & \textbf{3.07}\(\pm\).001 & \textbf{3.20}\(\pm\).002 & 6.27\(\pm\).01         & 11.78\(\pm\).01         \\
	\bottomrule
\end{tabular}

\end{table*}
Table~\ref{tab:avg_comp} presents the results averaged across all target datasets for the \gls{am} architecture.
Despite not introducing any additional parameter overhead, \gls{mpol} achieves competitive performance across both perceptual and signal-level metrics.
While \gls{mpol} is not able to match \gls{laden}'s exceptional \gls{pesq} performance, it approximately matches or exceeds \gls{laden} and RemixIT in all other metrics.
Fig.~\ref{fig:am_pesq_spider} reveals that \gls{mpol} yields a very consistent gain on the \gls{pesq} metric across all datasets, despite being on average smaller than \gls{laden}'s gain.
This consistency suggests that the confidence degradation is universal across domain shifts, constituting an important benefit over the more variable performance of \gls{laden}.
In combination with the low complexity, this consistency makes \gls{mpol} particularly suitable for edge deployments.
\par
The CMGAN results demonstrate \gls{mpol}'s effectiveness across different architectural paradigms.
Despite CMGAN's complex processing, including both masking and additive terms, \gls{mpol} achieves the highest average \gls{pesq} performance among all methods.
While \gls{mpol}'s improvements on other metrics are more modest compared to \gls{laden}, the method maintains competitive performance across all measures and outperforms RemixIT in all metrics.
This success on a more sophisticated architecture suggests that mask confidence degradation represents a fundamental phenomenon in \gls{tf} masking-based processing.
\par
These results, combined with the \gls{am} findings, confirm that mask polarization provides a robust adaptation signal across diverse \gls{se} architectures without requiring additional parameters.
\Gls{mpol}'s parity with \gls{laden} is particularly notable, given \gls{laden}'s 4M parameter encoder overhead.
\begin{figure}[h]
	\centering
	\resizebox{0.75\columnwidth}{!}{\begin{tikzpicture}
	\tkzKiviatDiagram[scale=0.9,label distance=0.0cm,
		gap=1.0,
		ymin=-0.15,
		ymax=0.15,
		label space=3.75cm,
        lattice=3]{{~~~~~~EARS-D},VBD,VBW,\dns{EN},{\dns{GE}\hspace{2em}~},{\dns{IT}\hspace{2em}~},\dns{RU},\dns{SP},\dns{FR}}
	\tkzKiviatLine[thick,color=mittelblau,mark=none,opacity=.5](
	0.0,
	0.0,
	0.0,
	0.0,
	0.0,
	0.0,
	0.0,
	0.0,
	0.0
	)
	\tkzKiviatLineError[thick,color=orange,opacity=.5](
	-0.007/0.006,
	-0.018/0.003,
	0.018/0.007,
	0.004/0.007,
	0.012/0.018,
	0.003/0.029,
	-0.001/0.005,
	0.005/0.007,
	0.009/0.030
	)
	\tkzKiviatLineError[thick,mark size=4pt,color =lila](
	-0.056/0.007,
	-0.081/0.006,
	-0.010/0.003,
	0.016/0.005,
	0.075/0.010,
	0.091/0.009,
	-0.002/0.004,
	0.046/0.006,
	0.102/0.012
	)

	\tkzKiviatLineError[thick,mark size=4pt,color=pink](
-0.008/0.001,
-0.025/0.001,
0.020/0.002,
0.059/0.004,
0.156/0.007,
0.166/0.004,
0.036/0.011,
0.097/0.007,
0.164/0.005
)
	\tkzKiviatGrad[prefix=,unity=1,suffix=](0)
\end{tikzpicture}}
	\resizebox{0.75\columnwidth}{!}{\begin{tikzpicture}
    \begin{axis}[%
            hide axis,
            xmin=10,
            xmax=50,
            ymin=0,
            ymax=0.4,
            legend style={font=\small, legend columns=-1},
            legend cell align={left},
        ]
        \addlegendimage{mittelblau, thick}
        \addlegendentry{Source \ }
        \addlegendimage{orange, thick}
        \addlegendentry{RemixIT \ }
        \addlegendimage{lila, thick}
        \addlegendentry{LaDen}
        \addlegendimage{pink, thick}
        \addlegendentry{MPol}
    \end{axis}
\end{tikzpicture}}
	\caption{\(\Delta\)PESQ results relative to the source performance using the CMGAN architecture (\(\mu\pm2\sigma\)).}
	\vspace{-1em}
	\label{fig:cmgan_pesq_spider}
\end{figure}
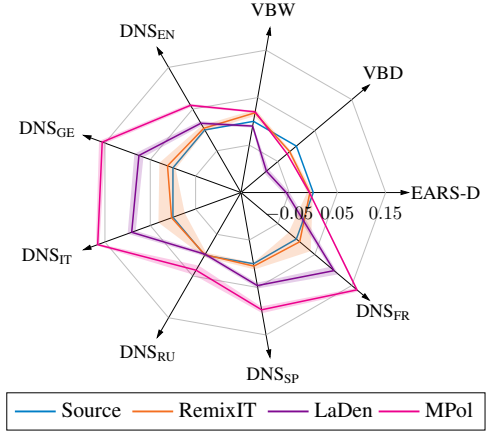
\vspace{-0.5em}
\subsection{Ablation Study}
The ablation study (Table~\ref{tab:ablation}) highlights the critical role of the weight ensembling.
While the Wasserstein loss \(\mathcal{L}_{\mathrm{W}}\) improves the perceptual performance on both datasets, it significantly degrades the signal-level performance.
The penalty term \(\mathcal{L}_{\mathrm{S}}\) mitigates the degraded signal-level performance but has a mixed effect on the perceptual performance.
Adding weight ensembling significantly improves the adaptation stability, leading to more consistent results across datasets and metrics.
\begin{table}[h]
	\centering
	\caption{Ablation study using the \gls{am} architecture.}
	\label{tab:ablation}
	\begin{tabular}{lcccc}
	\toprule
	                              & \multicolumn{2}{c}{VBD} & \multicolumn{2}{c}{\dns{GE}}                                           \\
	\cmidrule(r){2-3} \cmidrule(r){4-5}
	                              & PESQ \(\uparrow\)       & SI-SDR \(\uparrow\)          & PESQ \(\uparrow\) & SI-SDR \(\uparrow\) \\
	\midrule
	Source                        & 2.424                   & 11.487                       & 1.806             & 11.665              \\
	\midrule
	\(\mathcal{L}_{\mathrm{W}}\)  & 2.433                   & 10.462                       & 1.859             & 7.462               \\
	+\(\mathcal{L}_{\mathrm{S}}\) & 2.304                   & 10.811                       & \textbf{1.907}    & 9.367               \\
	+{\small Ensemble}            & \textbf{2.489}          & \textbf{11.741}              & 1.881             & \textbf{11.855}     \\
	\bottomrule
\end{tabular}

	\vspace{-1em}
\end{table}

\section{Conclusion}
We presented \gls{mpol}, a lightweight \gls{tta} method for speech enhancement that achieves competitive performance across diverse architectures without requiring any additional components.
Our key observation that mask-based \gls{se} models universally lose bimodal characteristics under domain shifts provides a natural adaptation signal that can be efficiently exploited through distribution-based polarization.
The method demonstrates that mask confidence restoration represents a robust approach to \gls{tta} for \gls{se}, enabling deployment in resource-constrained scenarios where existing methods' parameter overhead would be prohibitive.
This work establishes a foundation for practical \gls{tta} in speech enhancement by bridging insights from classification adaptation to generative audio tasks.


\bibliographystyle{IEEEbib}
\bibliography{refs}

\end{document}